\documentclass[final]{svjour3}
\usepackage{epstopdf}
\usepackage{amsmath}
\usepackage{amsfonts}
\usepackage{amssymb}
\usepackage{tabularx}
\usepackage{color}
\usepackage{multirow}
\usepackage{graphicx}
\smartqed  
\begin{document}
\sloppy
\newcommand{\beq}{\begin{equation}}
\newcommand{\eeq}{\end{equation}}
\newcommand{\divg}{\mbox{\rm{div}}\,}
\newcommand{\Divg}{\mbox{\rm{Div}}\,}
\newcommand{\D}  {\displaystyle}
\newcommand{\DS} {\displaystyle}
\def\sca   #1{\mbox{\rm{#1}}{}}
\def\mat   #1{\mbox{\bf #1}{}}
\def\vec   #1{\mbox{\boldmath $#1$}{}}
\def\scas  #1{\mbox{{\scriptsize{${\rm{#1}}$}}}{}}
\def\scat  #1{\mbox{{\footnotesize{${\rm{#1}}$}}}{}}
\def\scaf  #1{\mbox{{\tiny{${\rm{#1}}$}}}{}}
\def\vecs  #1{\mbox{\boldmath{\scriptsize{$#1$}}}{}}
\def\tens  #1{\mbox{\boldmath{\scriptsize{$#1$}}}{}}
\def\tenf  #1{\mbox{{\sffamily{\bfseries {#1}}}}}
\def\ten   #1{\mbox{\boldmath $#1$}{}}
\title{Bulging brains}
\author{J.~Weickenmeier \and P.~Saez \and A.~Goriely \and E.~Kuhl}
\institute{
	J. Weickenmeier \at
    Department of Mechanical Engineering, Stanford University, Stanford, CA 94305, USA,
	\email{weickenmeier@stanford.edu}           
\and
	P. Saez \at
	Laboratori de Calcul Numeric, Universitat Universitat Polit\`ecnica de Catalunya Barcelona\-Tech, 08034 Barcelona, Spain,
	\email{pablo.saez@upc.edu}           
\and
	A. Goriely \at
	Mathematical Institute, University of Oxford, Oxford, OX2 6GG, UK,
	\email{goriely@maths.ox.ac.uk}
\and
	E. Kuhl \at 
	Department of Mechanical Engineering and Department of Bioengineering, Stanford University, Stanford, CA 94305, USA,
	\email{ekuhl@stanford.edu}
}
\date{Received: date / Accepted: date}
\maketitle
\begin{abstract}
Brain swelling is a serious condition associated with an accumulation of fluid inside the brain that can be caused by trauma, stroke, infection, or tumors. It increases the pressure inside the skull and reduces blood and oxygen supply. To relieve the intracranial pressure, neurosurgeons remove part of the skull and allow the swollen brain to bulge outward, a procedure  known as decompressive craniectomy. Decompressive craniectomy has been preformed for more than a century; yet, its effects on the swollen brain remain poorly understood. Here we characterize the deformation, strain, and stretch in bulging brains using the nonlinear field theories of mechanics. Our study shows that even small swelling volumes of 28 to 56 ml induce maximum principal strains in excess of 30\%. For radially outward-pointing axons, we observe maximal normal stretches of 1.3 deep inside the bulge and maximal tangential stretches of 1.3 around the craniectomy edge. While the stretch magnitude varies with opening site and swelling region, our study suggests that the locations of maximum stretch are universally shared amongst all bulging brains. Our model can inform neurosurgeons and rationalize the shape and position of the skull opening, with the overall goal to reduce brain damage and improve the structural and functional outcomes of decompressive craniectomy in trauma patients.

\end{abstract}

\

\vspace*{0.1cm}
\noindent{\small{Einen \,Druck \"uber \,einen \,gewissen Grad und \"uber eine \,gewisse \,Zeit \,hinaus \,h\"alt \,das \hfill Gehirn \\[-2.pt]
nicht aus. Darum ist es stets Pflicht, \,einen Druck auf das ungef\"ahrliche \,Mass \,von \hfill Intensit\"at \\[-2.pt]
und Dauer zu reduciren. \hfill Emil Theodor Kocher [1901]}}
\section{Motivation}\label{Sec:1}
Under physiological conditions, the mechanical environment of our brain is tightly regulated. The intracranial pressure, for example, lies within a narrow window between 0 and 10\,mmHg \cite{goriely15}. An increase in intracranial pressure--most commonly caused by  traumatic brain injury, subarachnoid hemorrhage, ischemic stroke or a brain tumor--can be devastating or even fatal: It reduces cerebral perfusion, 
and limits the supply of metabolites and oxygen \cite{cooper11}. As a method of last resort, neurosurgeons remove part of the skull to allow the swollen brain to bulge outward and facilitate an immediate release of the elevated pressure \cite{kolias13}. This life-saving procedure, known as decompressive craniectomy \cite{quinn11}, is typically recommended if the intracranial pressure exceeds 20\,mmHg for longer than 30 minutes \cite{jones13}. While a decompressive craniectomy improves short-term pressure management and survival, its survivors often experience severe long-term disabilities \cite{hutchinson06}. To date, the precise criteria related to the optimal timing of treatment, the optimal location and size of the skull opening, and the long-term functional outcome remain unclear.

From a mechanical perspective, a decompressive craniectomy is a compromise between maximizing the management of the intracranial pressure and minimizing the deformations induced by the bulging brain \cite{goriely16}. Recent studies have characterized bulge kinematics based on computerized tomography images before and after a decompressive craniectomy using non-linear image registration \cite{holst12}; yet, little is know about the stress, stretch, and strain inside the brain. While our mechanical intuition tells us that stretch and strain can be reduced by increasing the opening size, a larger opening area is more prone to infection and clinically undesirable \cite{stiver09}. Clinical guidelines suggest opening diameters of about 12\,cm \cite{tagliaferri12}, but the rationale for this recommendation is rather vague and lacks a clear mechanistic understanding of the bulging process itself. 

When aiming to optimize the craniectomy size, it is unclear to which extent the location of the opening influences the stretch and strain profiles across the brain \cite{fletcher14a}. The most common procedure in clinical practice is a unilateral craniectomy with an opening on either the left or the right lateral skull depending on the side of the swelling \cite{quinn11}. Recent clinical studies have challenged the engineering intuition that a collateral craniectomy with the opening at the site of swelling is less invasive than a contralateral craniectomy with the opening at the opposite, non-injured side \cite{holst14}. While the opening size for a unilateral craniectomy is anatomically limited, a bilateral craniectomy with a bifrontal opening across both hemispheres provides sufficient anatomic space for large opening sizes \cite{kolias13}. Yet, the precise bulging kinematics for the different types of craniectomy are far from being completely understood. 

Mathematical models and computational simulations can provide analytical and numerical insight into the strain, stretch, and stress fields of bulging solids. Using the classic theory of contact mechanics \cite{barber83,johnson87}, we have recently shown that in the small deformation limit, the bulging problem is conceptually similar to an inverted punch problem \cite{goriely16}. This allows us to solve the bulging problem explicitly for a bulging half-space under plane strain, plane stress, and axisymmetric conditions. The explicit analytical solution for the stress field motivates the introduction of damage drops, drop-shaped zones of high and low shear stress with singularities that scale with the inverse square root of the distance from the opening \cite{weickenmeier16a}. Interestingly, the shape of the bulge, the singularities of the stress profile, and the orientation of the drops are generic for all bulging problems and independent of the constitutive model. These characteristic features also agree nicely with computational simulations, both in the linear \cite{fletcher14} and in the nonlinear \cite{weickenmeier16a} regime. However, it remains unclear how these characteristics evolve in geometries as complex as the human brain. First attempts along these lines have modeled the brain via its convex hull embedded in a rigid skull \cite{gao08}, and shown that shear strains can reach values up to 25\%, even for bulge volumes of only 22 ml \cite{fletcher16}. While these numbers clearly highlight the need for a kinematically and constitutively nonlinear formulation, the bulging brain has never been modeled using the nonlinear field theories of mechanics. 

Here we introduce a continuum model for bulging brains in the finite deformation setting. We model brain tissue as a swelling, elastically incompressible Mooney-Rivlin solid and illustrate how to translate its mathematical model into a general, nonlinear finite element environment. To illustrate the features of the bulging problem under finite deformations, we conduct a series of case studies and perform systematic sensitivity analyses with respect to the swelling area, the opening size, and the opening location. We then create a personalized brain model from magnetic resonance images and simulate two different cases of craniectomy, a left unilateral flap and a frontal flap. For both cases, we study three swelling scenarios, swelling in both hemispheres, exclusively in the left hemisphere, and exclusively in the right hemisphere. We report and compare displacements, deformations, radial and tangential stretches, and maximum principal strains. 
\section{Brain model}\label{sec:2}
To model brain tissue, we adopt a classical hyperelastic constitutive formulation \cite{holzapfel00}. We follow the recommendation to approximate  brain as an isotropic material since our deformation rates are moderate \cite{wright11}. To characterize the brain at finite deformations, we introduce the nonlinear deformation map $\vec{\varphi}$ and its gradient $\ten{F}=\nabla_X \vec{\varphi}$ with respect to the coordinates $\vec{X}$ in the undeformed reference configuration. 
We allow parts of the brain to swell \cite{lang14}, and decompose the deformation gradient multiplicatively into an elastic part $\ten{F}^{\scas{e}}$ and a swelling part $\ten{F}^{\scas{s}}$,
\beq
  \ten{F} 
= \nabla_X \vec{\varphi} 
= \ten{F}^{\scas{e}}
  \cdot \ten{F}^{\scas{s}}
  \qquad \mbox{with} \qquad
  J
= \mbox{det}(\ten{F})
= J^{\scas{e}} J^{\scas{s}}  \,.
\eeq
The Jacobian 
$J$ denotes the total volume change and
$J^{\scas{e}}=\mbox{det}(\ten{F}^{\scas{e}})$
and
$J^{\scas{s}}=\mbox{det}(\ten{F}^{\scas{s}})$
denote the volume change associated with the elastic deformation and with swelling. 
We then make two major kinematic assumptions: 
We assume that the elastic behavior is incompressible,
$J^{\scas{e}}=1$, such that the total volume change is caused exclusively by swelling,
$J=J^{\scas{s}}$, and that swelling is volumetric, 
$\ten{F}^{\scas{s}} = (J^{\scas{s}})^{1/3} \ten{I}$, such that the isochoric deformation is purely elastic 
$\bar{\ten{F}} = \ten{F}^{\scas{e}}$.
These assumptions imply that we can  decompose the  deformation gradient $\ten{F}$ into a volumetric contribution purely associated with swelling, $J=J^{\scas{s}}$, and an isochoric contribution purely associated with the elastic deformation, $\bar{\ten{F}}=\ten{F}^{\scas{e}}$,
\beq
  \ten{F} 
= \nabla_X \vec{\varphi} 
= J^{1/3} \bar{\ten{F}}
  \qquad \mbox{with} \qquad
  J=\mbox{det}(\ten{F})
  \qquad \mbox{and} \qquad
  \bar{\ten{F}}
= J^{-1/3} {\ten{F}}  \,.
\eeq
We introduce the left Cauchy-Green deformation tensor $\ten{b}$ and decompose it into its swelling-induced volumetric contribution in terms of the Jacobian $J$ and its elastic isochoric contribution $\bar{\ten{b}}$,
\beq
 \ten{b} 
=\ten{F} \cdot \ten{F}^{\scas{t}}
=J^{2/3} \bar{\ten{b}}
 \qquad \mbox{with} \qquad
 \bar{\ten{b}}
=\bar{\ten{F}} \cdot \bar{\ten{F}}^{\scas{t}} \,.
\eeq
To characterize the swelling-induced deformation, we explore three kinematic metrics associated with the Green-Lagrange strain tensor,
\beq
    \ten{E} 
=   \mbox{$\frac{1}{2}$} \, 
[\, \ten{F}^{\scas{t}} \cdot \ten{F} - \ten{I} \,] \,,
\eeq
the maximum principal strain, $\lambda_{E}^{\scas{max}}$, associated with the eigenvalue problem
of the Green-Lagrange strain tensor $\ten{E}$,
\beq
  \ten{E} \cdot \vec{n}_{E} = \lambda_{E} \, \vec{n}_{E}
  \quad \mbox{and} \quad
  \lambda_{E}^{\scas{max}} = \mbox{max} \{ \lambda_E \}\,,
\eeq
the normal stretch along the axon, 
and the tangential stretch perpendicular to the axon.
We then introduce the invariants $I_1$, $I_2$, and $I_3$, in terms of the left Cauchy-Green deformation tensor $\ten{b}$,
\beq
\begin{array}{llc@{\hspace*{0.5cm}}c@{\hspace*{0.5cm}}lll}
   I_1
&=&\mbox{tr} (\ten{b}) 
&& {\partial I_1}/{\partial\ten{b}}
&=& \ten{I} \\
   I_2 
&=&\frac{1}{2} [\, \mbox{tr}^2 (\ten{b})-\mbox{tr} (\ten{b}^2) \,]
&  \mbox{with}  
&  {\partial I_2}/{\partial\ten{b}}
&=&I_1\ten{I}-\ten{b} \\
   I_3 
&=&\mbox{det} (\ten{b})    
&& \partial J/\partial\ten{b}
&=&\frac{1}{2} \, J \, \ten{b}^{-1}
\end{array} 
\label{invariants}
\eeq
and their elastic, isochoric counterparts $\bar{I}_1$, $\bar{I}_2$, and $\bar{I}_3$, either in terms of the isochoric left Cauchy-Green deformation tensor $\bar{\ten{b}}$ or in terms of the isochoric principal stretches 
$\bar{\lambda}_1$, $\bar{\lambda}_2$, and $\bar{\lambda}_3$, 
\beq
\begin{array}{llclllll@{\hspace*{-0.05cm}}l@{\hspace*{-0.05cm}}l}
   \bar{I}_1 
&=&\mbox{tr} (\bar{\ten{b}}) 
&=&J^{-2/3} 
&  I_1
&= \bar{\lambda}_1^2 
&+ \bar{\lambda}_2^2 
&+ \bar{\lambda}_3^2 \\
   \bar{I}_2  
&=&\frac{1}{2} [ \, \mbox{tr}^2 (\bar{\ten{b}})-\mbox{tr} (\bar{\ten{b}}^2) \, ]
&=&J^{-4/3} 
&  I_2    
&= \bar{\lambda}_1^{-2} 
&+ \bar{\lambda}_2^{-2} 
&+ \bar{\lambda}_3^{-2} \\
   \bar{I}_3 
&=&\mbox{det} (\bar{\ten{b}})
&=&J^{-6/3} 
&  I_3     
&= \, 1 \,.
\end{array} 
\label{iso_invariants}
\eeq
Many common constitutive models for brain tissues are special cases of the general Ogden model \cite{ogden72},
\beq
  \bar{\psi} 
= \sum_{\scas{i}=1}^{N}
  \frac{c_{\scas{i}}}{\alpha_i} 
  \left[\, \bar{\lambda}_1^{\alpha_{\scas{i}}} 
         + \bar{\lambda}_2^{\alpha_{\scas{i}}}
         + \bar{\lambda}_3^{\alpha_{\scas{i}}} -3 \, \right] \, ,
\label{ogden}
\eeq
parameterized in terms of  the Ogden parameters $c_{\scas{i}}$ and $\alpha_{\scas{i}}$. For the special case of $N=2$, with $\alpha_1=2$ and $\alpha_2=-2$, the Ogden model simplifies to the Mooney-Rivlin model \cite{mooney40,rivlin48},
\beq
  \bar{\psi} 
= \mbox{$\frac{1}{2}$} \,
  c_1 \, [\,\bar{\lambda}_1^2 + \bar{\lambda}_2^2 + \bar{\lambda}_3^2  - 3 \,] 
+ \mbox{$\frac{1}{2}$} \,
  c_2 \, [\,\bar{\lambda}_1^{-2} + \bar{\lambda}_2^{-2} + \bar{\lambda}_3^{-2} - 3 \,] \, .
\label{mooney}
\eeq
which we can reformulate in terms of the elastic isochoric invariants 
$\bar{I}_1$ and $\bar{I}_2$, 
\beq
  \bar{\psi} 
= \mbox{$\frac{1}{2}$} \,
  c_1 \, [\, \bar{I}_1 - 3 \,] 
+ \mbox{$\frac{1}{2}$} \,
  c_2 \, [\, \bar{I}_2 - 3 \,]\,.
\label{mooney_inv01}
\eeq
The  Mooney-Rivlin parameters $c_1$ and $c_2$ are related to the  shear modulus $\mu$ as $c_1 + c_2 = \frac{1}{2}\mu$, and their values can be identified through finite deformation experiments \cite{franceschini06,mihai15}. We enforce the elastic instability constraint, $J^{\scas{e}} - 1 = 0$, in the form, $J - J^{\scas{s}} = 0$, via a Lagrange multiplier $p$, and add the constraint $p \, [J-J^{\scas{s}}]$ to the energy functional, 
\beq
  \psi 
= \mbox{$\frac{1}{2}$} \,
  c_1 \, [\, \bar{I}_1 - 3 \,] 
+ \mbox{$\frac{1}{2}$} \,
  c_2 \, [\, \bar{I}_2 - 3 \,]
+ p \, [J-J^{\scas{s}}]  \,.
\label{mooney_inv02}
\eeq
To derive the stresses, it proves convenient to reformulate the energy in terms of the overall invariants $I_1$ and $I_2$ and the Jacobian $J$,
\beq
  \psi 
= \mbox{$\frac{1}{2}$} \,
  c_1 \, [\, J^{-2/3} \, I_1 - 3 \,] 
+ \mbox{$\frac{1}{2}$} \,
  c_2 \, [\, J^{-4/3} \, I_2 - 3 \,]
+ p \, [J-J^{\scas{s}}]  \,.
\label{psi_all}
\eeq
We can then directly obtain the Kirchhoff stress, 
\beq
  \ten{\tau}
= \frac{\partial\psi}{\partial\ten{F}} \cdot \ten{F}^{\scas{t}}  
= 2 \frac{\partial\psi}{\partial\ten{b}}  \cdot \ten{b}
= 2 \left[ 
    \frac{\partial\psi}{\partial I_1}
    \frac{\partial I_1}{\partial\ten{b}} 
+   \frac{\partial\psi}{\partial I_2} 
    \frac{\partial I_2}{\partial\ten{b}} 
+   \frac{\partial\psi}{\partial J} 
    \frac{\partial J}{\partial\ten{b}}  
    \right] \cdot \ten{b}
\label{kirchhoff01}
\eeq
or, with the derivatives of the invariants in Equation (\ref{invariants}), 
\beq
  \ten{\tau}
= 2 \left[ 
    \frac{\partial\psi}{\partial I_1} + I_1 \frac{\partial\psi}{\partial I_2}
    \right] \ten{b}
+ 2 \frac{\partial\psi}{\partial I_2} 
    \ten{b}^2 
+   J \frac{\partial\psi}{\partial J} 
    \ten{I}\,.
\label{kirchhoff02}
\eeq
Using the definition of the energy (\ref{psi_all}), we obtain the following explicit representation of the Kirchhoff stress $\ten{\tau}$ for a volumetrically swelling, elastically incompressible, Mooney-Rivlin material \cite{goriely15a},
\beq
  \ten{\tau}
= [\,  c_1  + \bar{I}_1 c_2 \, ] \, \bar{\ten{b}}   
- c_2 \, \bar{\ten{b}}^2
-[\, \mbox{$\frac{1}{3}$} \bar{I}_1 c_1 
+ \mbox{$\frac{2}{3}$} \bar{I}_2 c_2 + J p \,] \, \ten{I} \,.
\label{kirchhoff03}
\eeq
The isochoric contributions to the third term,  
$ \mbox{$\frac{1}{3}$} \bar{I}_1 c_1 
+ \mbox{$\frac{2}{3}$} \bar{I}_2 c_2$, 
reflect the fact that we have formulated the Mooney-Rivlin model in terms of the isochoric invariants $\bar{I}_1$ and $\bar{I}_2$ and not of the total invariants $I_1$ and $I_2$. Even though the elastic behavior is incompressible, the overall behavior is not, and the isochoric invariants $\bar{I}_1$ and $\bar{I}_2$ indirectly depend on the amount of swelling $J$. Rather than rewriting the energy formulation in Equation (\ref{psi_all}), we could have introduced the Kirchhoff stress as 
$ \ten{\tau} 
= \partial \Psi / \partial \bar{\ten{b}} 
: \mathbb{P} \cdot \ten{b}$, 
where $\mathbb{P} =\partial \bar{\ten{b}} / \partial \ten{b}$
denotes the spatial fourth order isochoric projection tensor, to obtain the term,
$ \mbox{$\frac{1}{3}$} \bar{I}_1 c_1 
+ \mbox{$\frac{2}{3}$} \bar{I}_2 c_2$,
from the isochoric projection with $\mathbb{P}$ \cite{holzapfel00}.

In our continuum model, we prescribe the amount of swelling $J^{\scas{s}}$ pointwise and phenomenologically rather than modeling the swelling process itself \cite{lang14}. We gradually increase the local tissue volume as 
$\Delta V = [\, J^{\scas{s}} -1.0 \,] \cdot 100\%$.
In our computational model, we represent volumetric swelling via volumetric thermal expansion \cite{abaqus14}, and only allow selected regions of the cerebral white matter tissue to swell, while all other substructures remain purely elastic with $J^{\scas{s}} \doteq 1.0$.  
We enforce the incompressibility constraint,
$p\, [J-J^{\scas{s}}]$, by using a hybrid finite element formulation with displacement degrees of freedom for the isochoric part and pressure degrees of freedom for the volumetric part of the deformation.  
\section{Bulging of a hemidisk}\label{sec:3}
Our previous analysis of the bulging of a linear elastic half-space through an opening has revealed two interesting features  related to damage and stress distributions and relevant to the problem of craniectomy: Large fiber stretches develop deep in the center of the bulge and large shear stresses develop around the opening edge. These results were obtained under the assumption of uniform swelling in a rectangular half-space geometry. In this section, we study the importance of geometric effects in an idealized geometry and systematically varying the location and area of swelling.
\subsection{Hemidisk model}
\begin{figure}[h]
\centering
\includegraphics[width=0.8\linewidth]{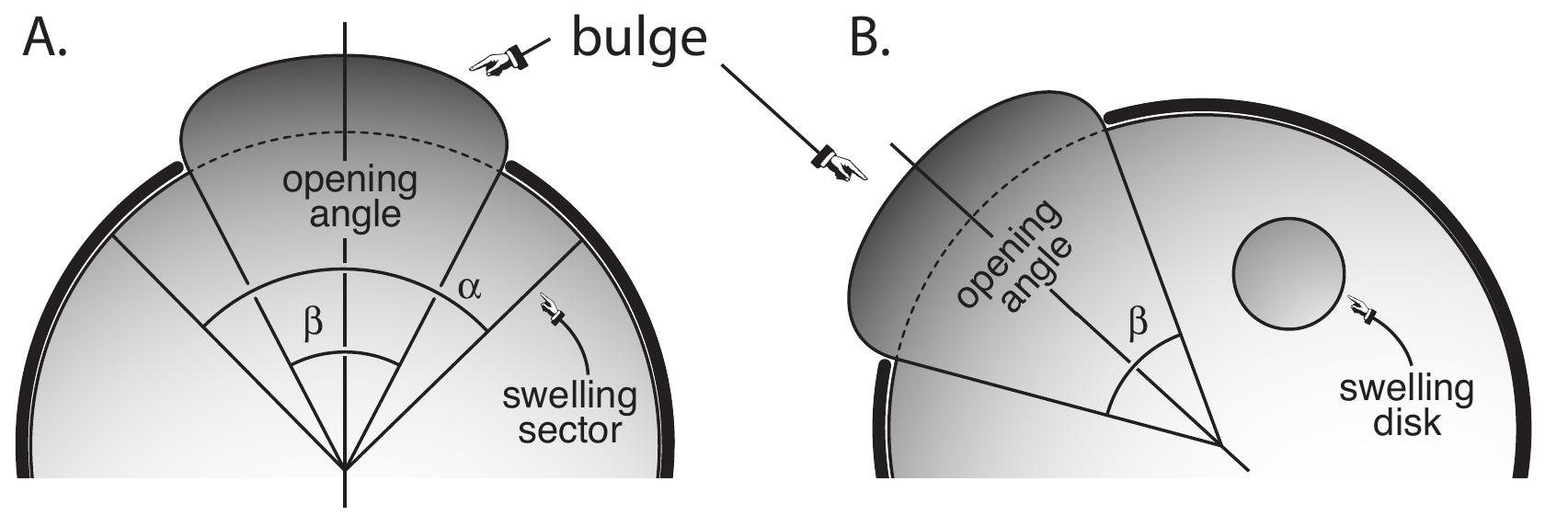}
\caption{Bulging of a hemidisk. We allow an elastic body to swell locally, either in a sector of angle $\alpha$ (A) or in a disk (B). The swelling body bulges out through an opening of angle $\beta$.}
\label{fig01}
\end{figure}

We first consider the bulging problem in a simple two-dimensional geometry. As depicted in Figure~\ref{fig01}, an incompressible isotropic elastic hemidisk is swelling and the deformations are constrained within the hemidisk except in an opening of angle $\beta$. We consider two swelling scenarios: the swelling of a sector with an opening $\alpha$ where both $\alpha$ and $\beta$ are centered about the axis of symmetry as illustrated in Figure~\ref{fig01}A, and the swelling of a disk where the opening $\beta$ is inclined off the axis of symmetry as illustrated in Figure~\ref{fig01}B. We assume that the axonal direction $\vec{n}$ is oriented radially outward, with $\vec{t}$ denoting the tangential direction. For both swelling scenarios, we present the radial or normal stretch 
$\lambda_n
= [\, \vec{n} \cdot \ten{F}^{\scas{t}}
   \cdot \ten{F} \cdot \vec{n} \,]^{1/2}$
and the tangential or shear stretch
$\lambda_t
= [\, \vec{t} \cdot \ten{F}^{\scas{t}}
   \cdot \ten{F} \cdot \vec{t} \,]^{1/2}$.
\subsection{Bulging of a hemidisk with a swelling sector}
An important consideration for the stress distribution within the solid is the type of contact. By definition, the boundary of the bulge is traction free. On the base of the hemidisk, we assume no sliding. On the curved part of the contact region, we use two types of boundary conditions: either frictional contact without sliding or frictionless contact with sliding. To visualize both contact conditions side by side, we only show half of the hemidisk for each contact condition, frictional contact on the left and frictionless contact on the right. 

\begin{figure}[h]
\centering
\includegraphics[width=\linewidth]{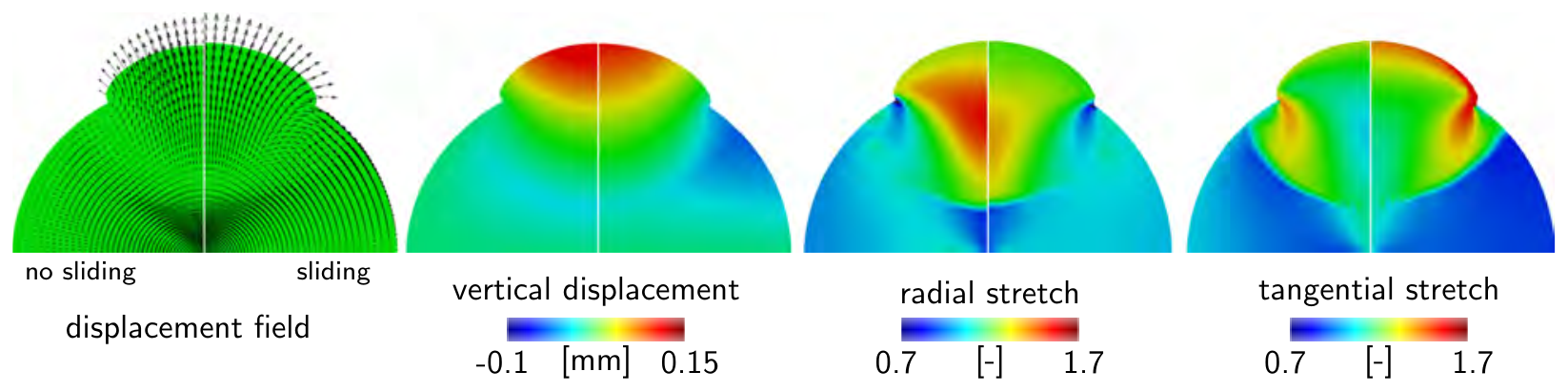}
\caption{Bulging of a hemidisk with a swelling sector.  
Displacement, vertical displacement, and radial and tangential stretches for frictional contact without sliding (left) and frictionless contact with sliding (right). In frictional contact without sliding, the solid is pushed outward with large displacements along the symmetry axis in the center of the bulge. In frictionless contact with sliding, the solid slides along the boundary and rotates outward around the opening edge.}
\label{fig02}
\end{figure}
Figure \ref{fig02} illustrates the impact of the contact condition for a swelling sector of $\alpha=80^0$ and an opening angle  of $\beta=60^0$ at a swelling magnitude of of $J=1.2$. For frictional contact without sliding shown on the left, the boundary nodes are fixed. Upon swelling, the solid is pushed outward with large displacements along the symmetry axis in the center of the bulge.  For frictionless contact with sliding, the boundary nodes are allowed to slide freely along the contact region. Upon swelling, the solid slides along the boundary and rotates outward around the opening edge. 

\begin{figure}[h]
\centering
\includegraphics[width=\linewidth]{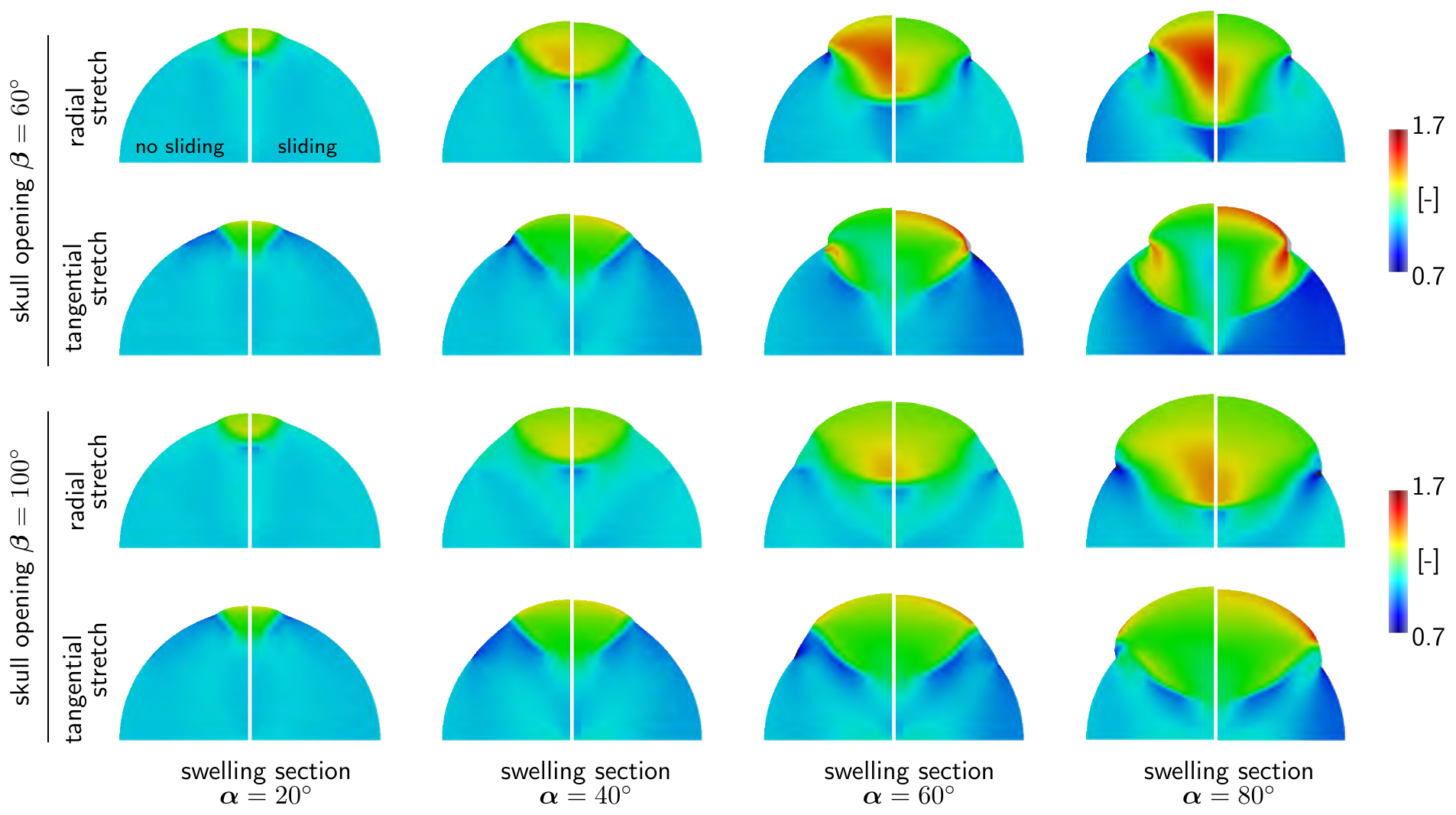}
\caption{Bulging of a hemidisk with a swelling sector. 
Radial and tangential stretches for varying opening angles of $\beta$ (rows) and for varying swelling sector angles $\alpha$ (columns) for frictional contact without sliding (left) and frictionless contact with sliding (right). Radial stretches take maximum values of 1.7 in regions deep inside the bulge; tangential stretches take maximum values of 1.7 in regions localized around the craniectomy edge.}
\label{fig03}
\end{figure}

Figure~\ref{fig03} illustrates a sensitivity analysis with respect to the opening angle $\beta$, the swelling sector angle $\alpha$, and the contact condition for a swelling magnitude of of $J=1.2$. As the angle  $\alpha$ of the swelling sector increases with $\alpha=20^0, 40^0, 60^0, 80^0$, from left to right, the relative swelling area increases as 
$\Delta A/A=\alpha / \pi [\, J-1 \,]$. Figure~\ref{fig03} reveals a number of expected and new features:
(i) As expected, an increase in skull opening reduces the maximal deformation and with it the maximal stretch;
(ii) The radial stretch is maximal in a zone deep inside the bulge and increases rapidly as the swelling increases;
(iii) The tangential stretch is maximal  at the opening edge in a zone that takes the form of a drop;
(iv) The maximal radial stretch is markedly higher in the case of frictional contact without sliding than in the frictionless contact case;
(v) The maximal tangential stretch is markedly higher in the case of frictionless contact with sliding than in the frictional contact case.
These features and trends appear to be shared broadly by all bulging cases.
\subsection{Bulging of a hemidisk with a swelling disk}
In the case of tumor-induced swelling, it is likely that the swelling region takes a spherical rather than a sector shape. To explore the effects of a swelling disk and analyze the sensitivity of the swelling location with respect to the location of the skull opening, we study five cases with varying swelling locations for a swelling of $J=1.3$. For all five cases, we model the contact region as frictionless with sliding.

\begin{figure}[h]
\centering
\includegraphics[width=\linewidth]{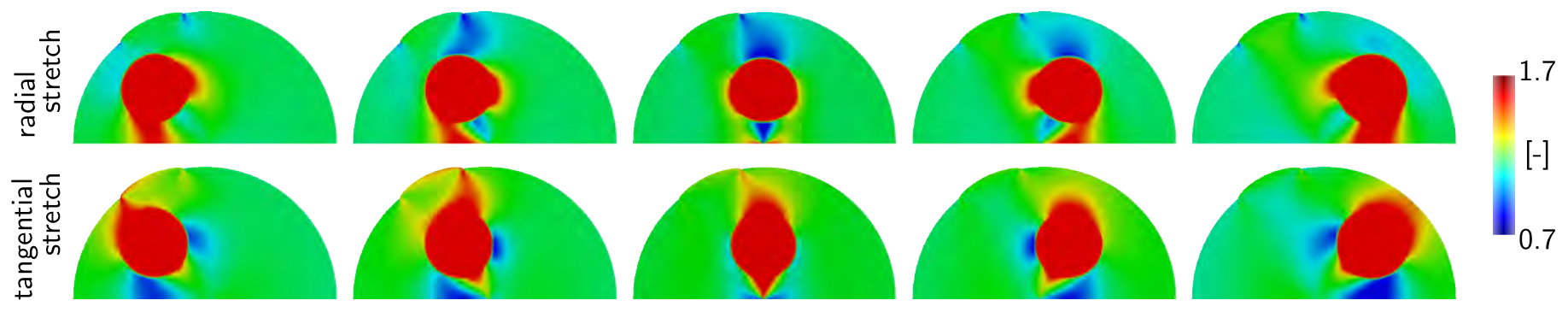}
\caption{Bulging of a hemidisk with a swelling disk. 
Radial and tangential stretches (rows) for five different swelling locations (columns). Radial and tangential stretches take maximal and mimimal values around the swelling disk, while large regions of the hemidisk are unaffected by the local swelling.}
\label{fig04}
\end{figure}
Figure~\ref{fig04} illustrates the radial and tangential stretches for the bulging hemidisk with a varying position of the swelling disk. In all five cases, swelling is a local event. Maximal and minimal stretches are localized close to the swelling disk. Except for the swelling region itself, the overall stretch profile is rather insensitive to the location of swelling. 
\section{Bulging of a personalized brain}\label{sec:4}
To simulate the effects of swelling in an anatomically detailed brain geometry, we create a personalized human head model from magnetic resonance images and simulate six different scenarios: a decompressive craniectomy with either unilateral flap or frontal flap subjected to both left and right, exclusively left, and exclusively right hemispherical swelling of the white matter tissue.
\subsection{Personalized brain model}\label{sec:41}
\begin{figure}
\centering
\includegraphics[width=\linewidth]{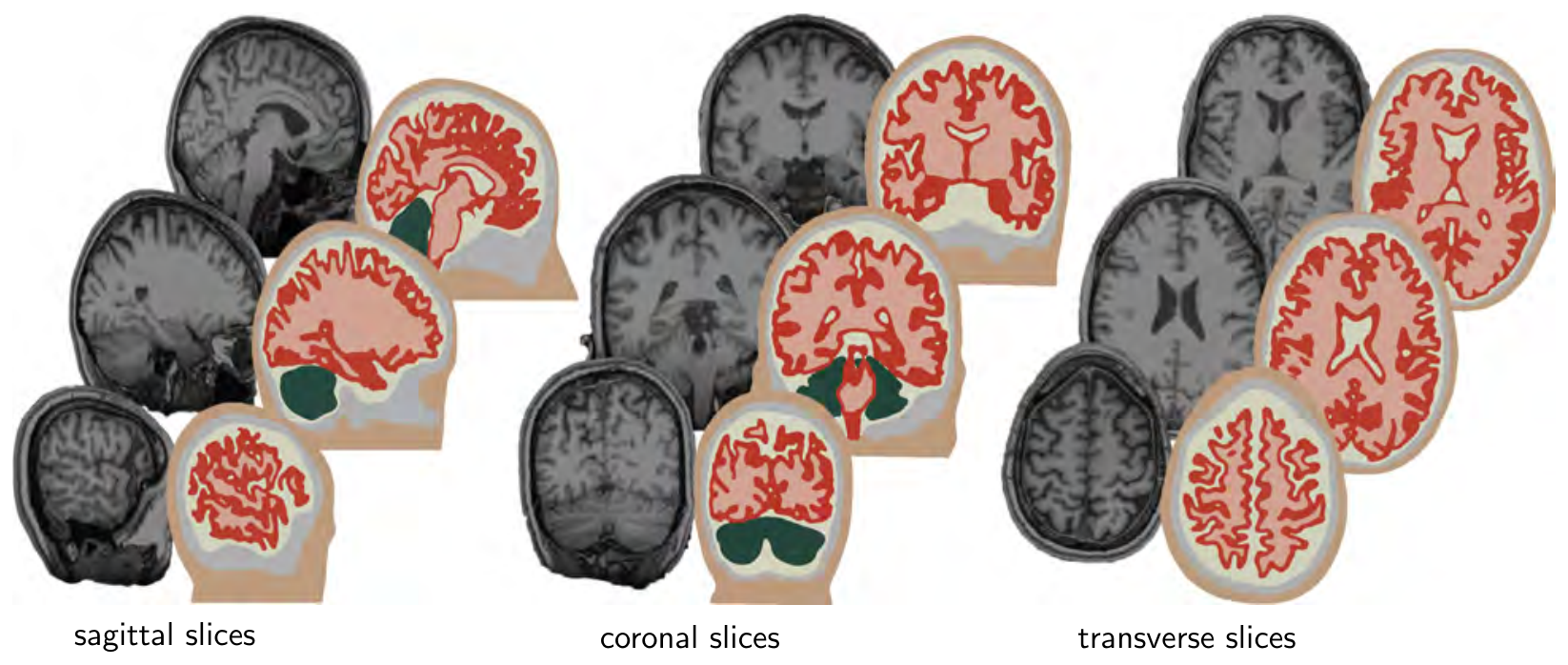}
\caption{Personalized decompressive craniectomy model. 
Magnetic resonance images (left) and computational model (right).
Anatomically detailed and geometrically accurate three-dimensional reconstructions of the individual substructures including the gray matter (red), the white matter (pink), the cerebrospinal fluid (beige), the cerebellum (green), the skin (brown), and the skull (gray) shown for selected sagittal, coronal, and transverse slices.}
\label{fig05}
\end{figure}

Figure \ref{fig05} shows representative sagittal, coronal, and transverse slices of an adult female head that form the basis of our anatomic model. The brain has a total volume of 1,108cm$^3$, a surface area of 1,673cm$^2$, and an average cortical thickness of 0.252mm. Our magnetic resonance image set contains a total of 190 slices in the sagittal plane at a spacing of 0.9 mm. Each slice has a matrix representation of 256 $\times$ 256 pixels with an in-plane resolution of 0.9 mm $\times$ 0.9 mm \cite{saggar15}. From the magnetic resonance images, we create a personalized high-resolution anatomic model of the brain using the ScanIP software environment of Simpleware \cite{young08}. This semi-automatic software iteratively produces an anatomically detailed and geometrically accurate three-dimensional reconstruction of all relevant substructures including the cerebral gray and white matter, the cerebrospinal fluid, the cerebellum, the skin, and the skull \cite{cotton16}. From these substructures, we create a finite element model with 1,275,808 linear tetrahedral elements and 241,845 nodes using the finite element meshing tool of Simpleware \cite{weickenmeier16b}. We import our head model into the finite element software package Abaqus, in which we prescribe the constitutive models as well as  the boundary, contact, and loading conditions \cite{abaqus14}. 

\begin{table}[h]
\centering
\renewcommand{\arraystretch}{1.1}
\caption{Material parameters of the Mooney-Rivlin model in different regions of the brain.}
\label{tab01}

\vspace*{0.2cm}
\begin{tabular}{|p{0.25\linewidth}| c | c |}
\hline
\textbf{\textsf{substructure}} & 
\textbf{\textsf{parameter} $c_1$ \textsf{[kPa]}} & 
\textbf{\textsf{parameter} $c_2$ \textsf{[kPa]}}\\ \hline \hline
cerebral gray matter & 0.28 & 333.0 \\
cerebral white matter & 0.56 & 666.0  \\ 
cerebellum & 0.28 & 333.0\\ 
cerebrospinal fluid & 0.03 & $\,\,\,$33.3 \\\hline
\end{tabular}
\end{table}
For the constitutive model, we adapt a Mooney-Rivlin model with gray matter parameters $c_1=0.28$\,kPa and $c_2=333$\,kPa \cite{mihai15}. We assume that the cerebellum is as stiff as the gray matter tissue, and that the white matter tissue is twice as stiff \cite{budday15}. For simplicity, we model the cerebrospinal fluid as an ultrasoft solid with a stiffness ten times lower  than the gray matter stiffness. We assume that all soft tissues are incompressible and enforce the incompressibility constraint using hybrid linear tetrahedral C3D4H elements \cite{abaqus14}. Table \ref{tab01} summarizes our material parameters for the individual substructures of the brain.

\begin{figure}[h]
\centering
\includegraphics[width=0.9\linewidth]{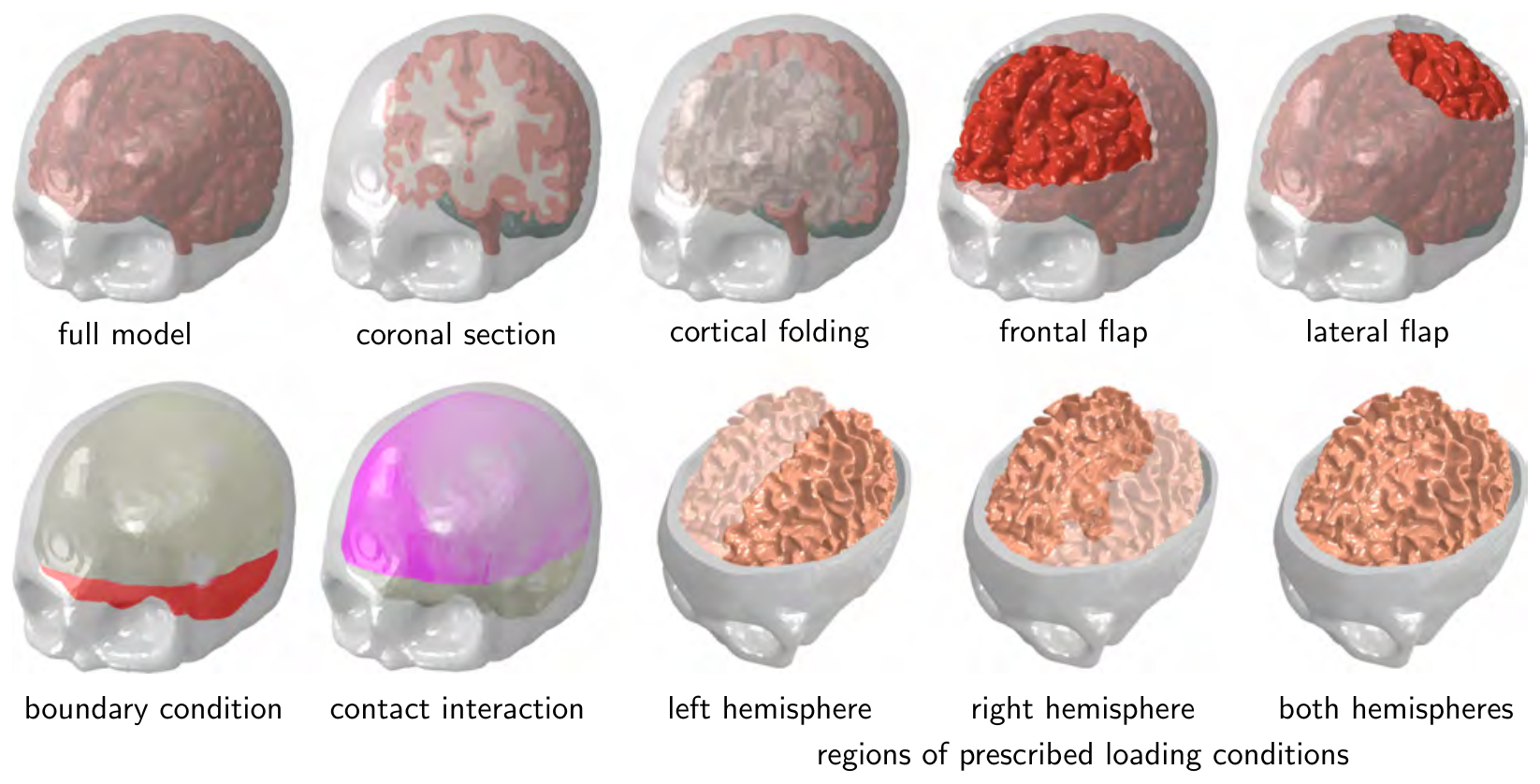}
\caption{Personalized decompressive craniectomy model. Boundary conditions and loading conditions. Top row: Full model discretized with 1,275,808 linear tetrahedral elements and 241,845 nodes; representative coronal section; anatomic details with cortical folds; frontal flap with 4,279 skull elements removed; lateral flap with 2,494 elements removed. Bottom row: Boundary conditions with inferior brain tissues and  superior cerebral spinal fluid fixed relative to the skull (red) and remaining outer brain surface allowed to slide along the inner skull (pink); swelling of left, right, and both white matter hemispheres.}
\label{fig06}
\end{figure}
For the boundary conditions, we use combinations of fixed and sliding contact at the outer brain surface. Figure \ref{fig06} illustrates our boundary conditions across the brain. To limit the motion of the inferior soft tissue regions, we apply homogeneous Dirichlet boundary conditions, shown in red, at the lower outer surface of the cerebrospinal fluid \cite{fletcher16}. To reduce the computational time, we ignore the skin layer and model the skull as a rigid body. We assume a tight contact between gray and white matter, the cerebellum, and the cerebrospinal fluid \cite{fletcher16}. At the interface between the cerebrospinal fluid and the skull, we apply frictionless contact, shown in pink, to  allow the brain to slide freely along the skull \cite{weickenmeier16b}. 

For the loading conditions, we simulate brain swelling by prescribing a local volumetric expansion in a predefined white matter region. Figure \ref{fig06} illustrates our loading conditions. We gradually increase the amount of swelling from $J^{\scas{s}}=1.0$ to $J^{\scas{s}}=1.1$ to model a volumetric expansion of 10\% in selected regions of the white matter tissue. 
\subsection{Bulging of a personalized brain with swelling white matter tissue}
We simulate three different cases of swelling, in both hemispheres, exclusively in the left hemisphere, and exclusively in the  right hemisphere. To release the swelling-induced pressure, we simulate two different decompressive craniectomies, a frontal flap with 4,279 skull elements removed and a unilateral flap with 2,494 elements removed. For all cases, we quantify and compare the mechanical response in terms of the overall deformation, the maximum principal strain, the radial and tangental stretch, and the midline shift. The midline shift is a common clinical indicator to characterize the degree of subcortical swelling and axonal damage. 

\begin{figure}
\centering
\includegraphics[width=\linewidth]{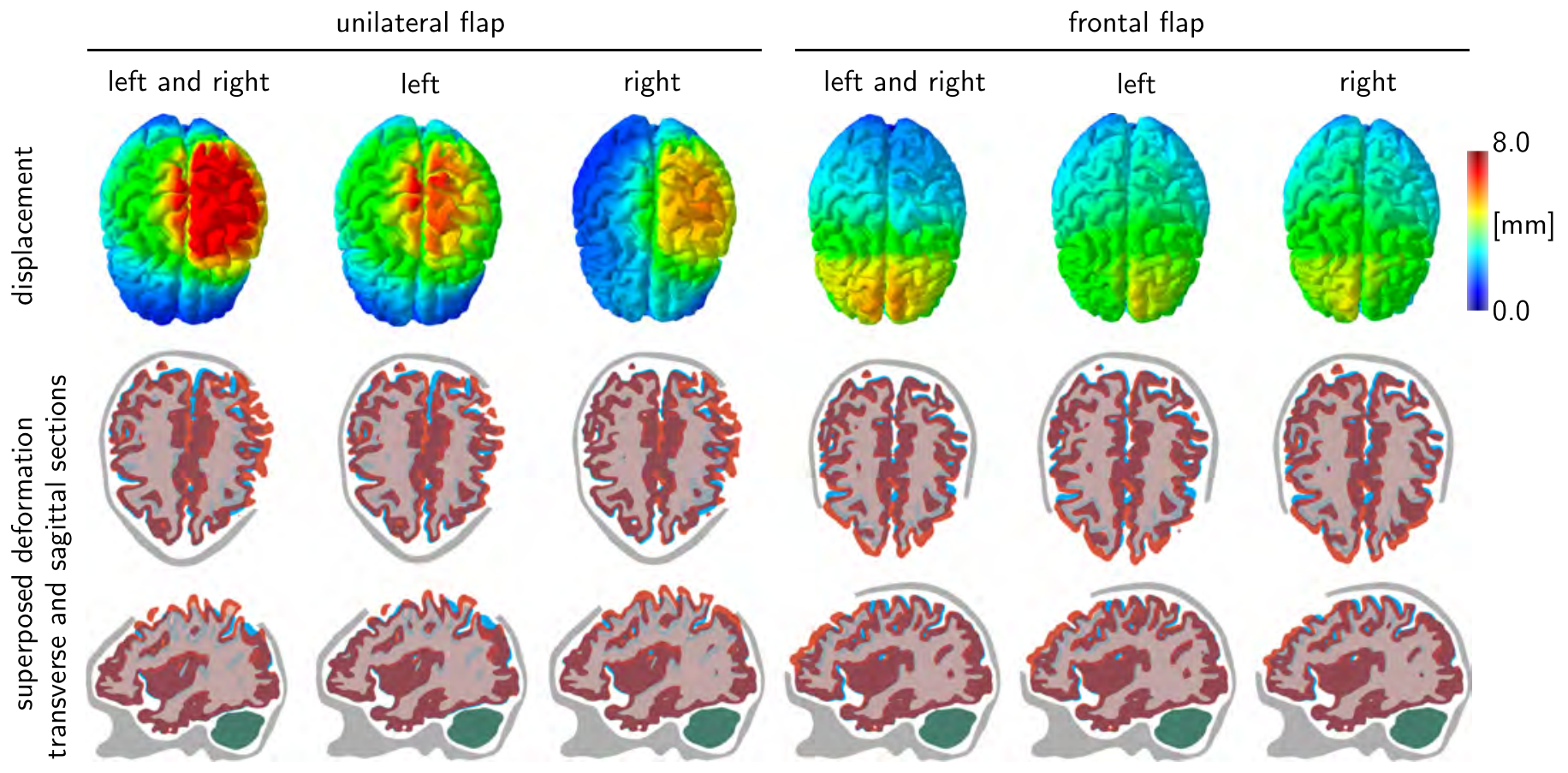}
\caption{Decompressive craniectomy. Displacement and superposed deformation in transverse and sagittal sections for unilateral and frontal flaps with left and right, left, and right hemispherical swelling. Swelling causes a shift of intracranial tissues, a key indicator of the trauma's severity in clinical practice. The midline shift of the cortical and subcortical layers highlights the immediate release of tissue strain upon removal of the unilateral and frontal flaps.}
\label{fig07}
\end{figure}
Figure \ref{fig07} illustrates the displacement and the superposed deformation in transverse and sagittal sections for unilateral and frontal flaps with both left and right, only left, and only right hemispherical swelling. The surgical area available for a frontal flap is about twice as large as the area for a unilateral flap. Consequently, for the same amount of swelling, the displacements of the frontal flap are significantly smaller than for the unilateral flap. This finding is in agreement with our intuition and with our idealized hemidisk simulation in Figure \ref{fig03}, for which larger opening angles generate smaller radial and tangential stretches. The superposed deformation in transverse and sagittal sections in Figure \ref{fig07} highlights the relative motion of different regions of the brain as the brain bulges outward. Swelling naturally causes  a shift of all intracranial tissues. The shift of the midline, which is clearly visible in this sequence of images, is a key clinical indicator for the degree of trauma. 

\begin{figure}
\centering
\includegraphics[width=\linewidth]{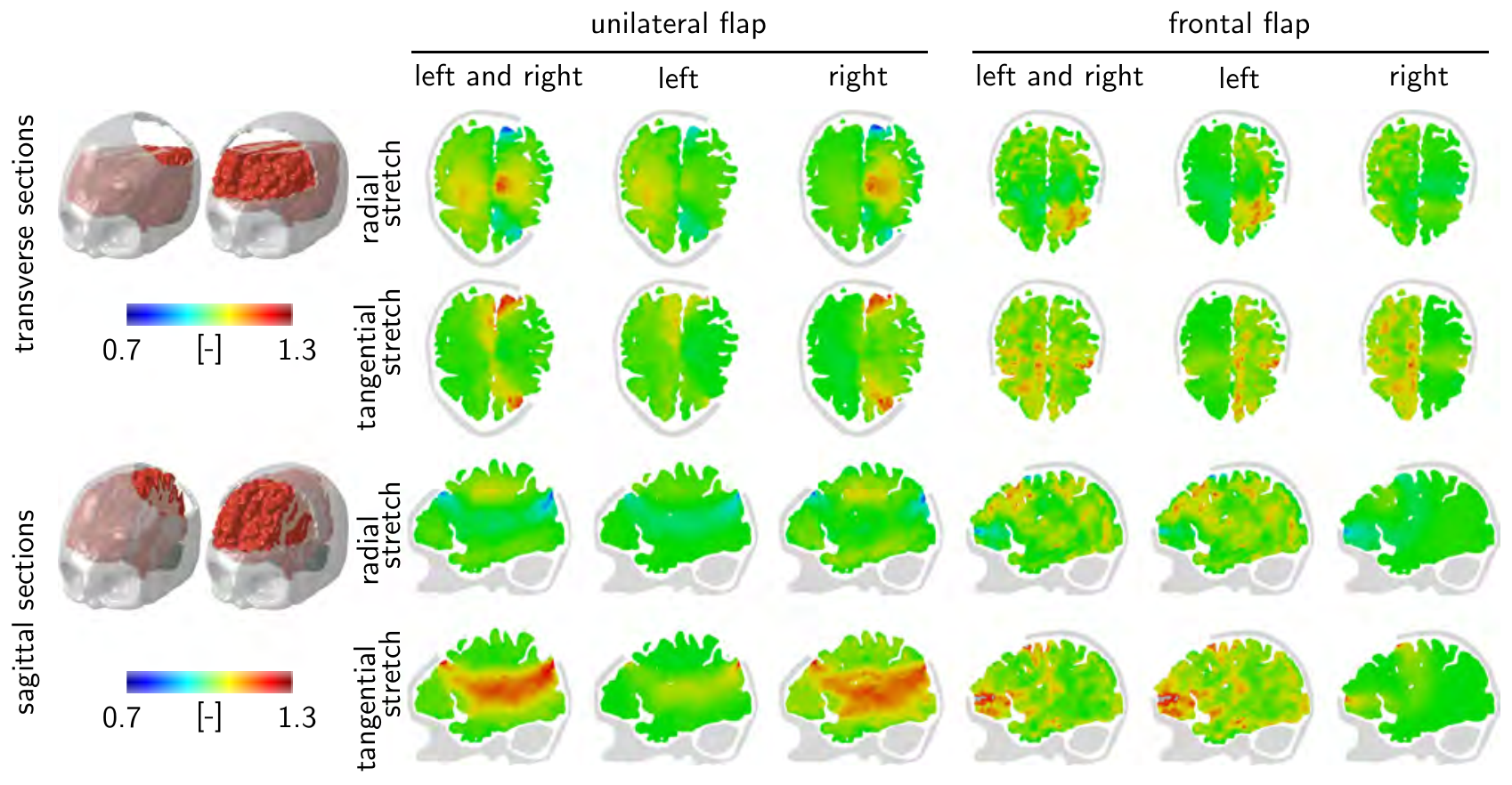}
\caption{Decompressive craniectomy. Radial and tangential stretches in transverse and sagittal sections for unilateral and frontal flaps with left and right, left, and right hemispherical swelling. 
Swelling causes maximum radial stretches of up to 1.3 deep inside the bulge, minimum radial stretches of 0.7 around the opening, and maximum tangential stretches of up to 1.3 around the opening.}
\label{fig08}
\end{figure}

Figure \ref{fig08} illustrates the radial and tangential stretches in transverse and sagittal sections for unilateral and frontal flaps with both left and right, only left, and only right hemispherical swelling. If we assume that axons are primarily oriented outward, we can associate the radial stretch with the axonal stretch and the tangential stretch with the axonal shear. For a swelling of 10\%, the radial stretch takes maximal values of up to 1.3 deep inside the bulge and minimal values of 0.7 around the edge of the opening. The tangential stretch takes maximal values of up to 1.3 in a ring around the opening. These three regions might be associated with potential zones of herniation and axonal failure, either by tension or compression, or by shear. 

\begin{figure}
\centering
\includegraphics[width=\linewidth]{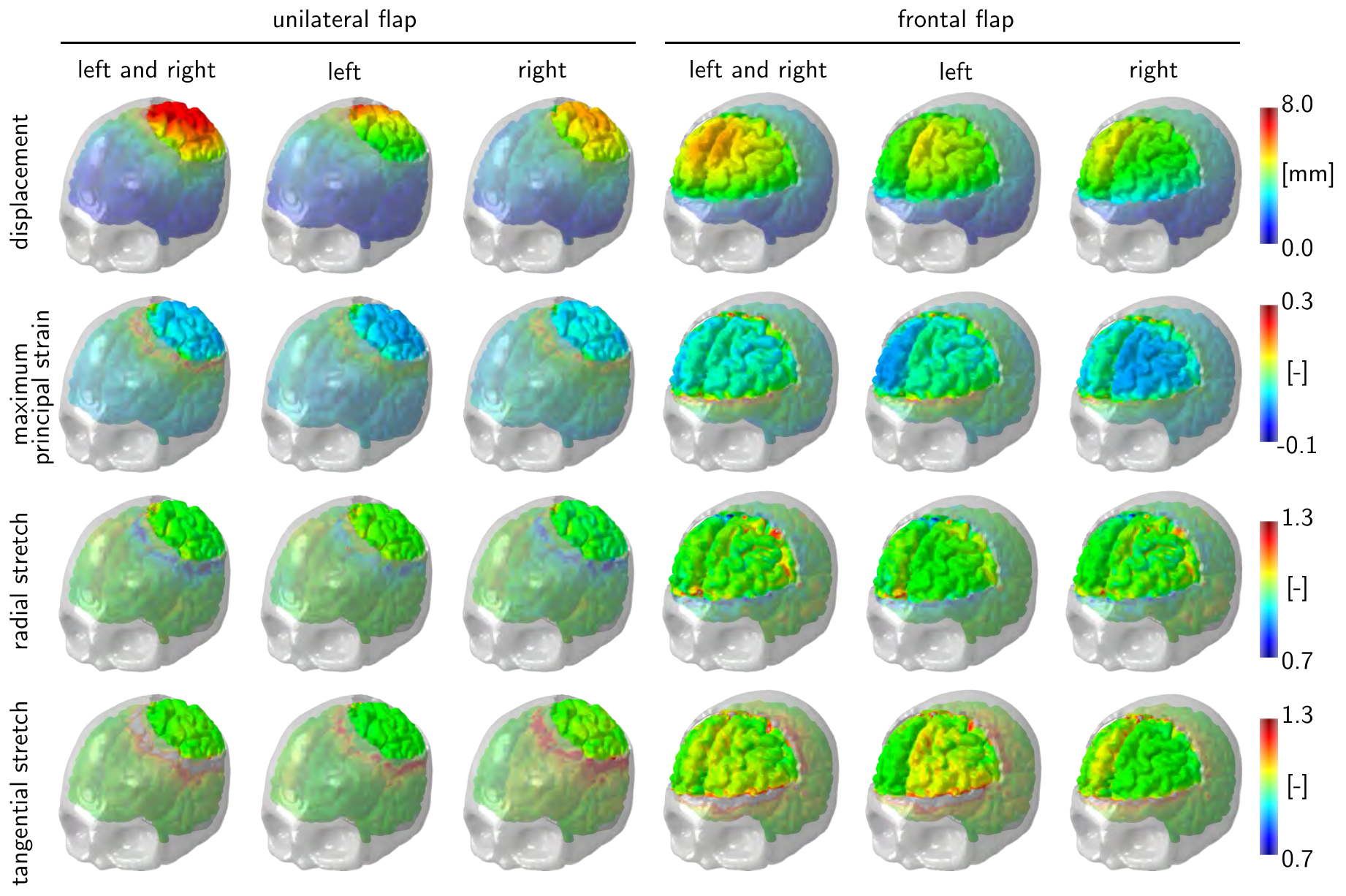}
\caption{Decompressive craniectomy. Displacement, maximum principal strain, radial stretch, and tangential stretch for unilateral and frontal flaps with left and right, left, and right hemispherical swelling. Swelling causes maximum principal strains of up to 30\% localized around the opening, maximum radial stretches of up to 1.3 deep inside the bulge, minimum radial stretches of 0.7 around the opening, and maximum tangential stretches of up to 1.3 around the opening.}
\label{fig09}
\end{figure}

Figure \ref{fig09} illustrates the displacement, maximum principal strain, radial stretch, and tangential stretch for unilateral and frontal flaps with both left and right, only left, and only right hemispherical swelling.  The displacement field confirms that the deformation is larger for the unilateral flap than for the frontal flap. Naturally, the displacements are largest in the center of the bulge, which explains the large radial strains in the bulge region. In agreement with Figure \ref{fig08}, the radial stretch takes maximum values of 1.3 deep inside the bulge and minimum values of 0.7 around the opening. Similarly, the tangential stretch takes maximum values of 1.3 in a ring around the opening. 

\section{Discussion}\label{sec:5}
Decompressive craniectomy is an invasive neurosurgical procedure to release elevated pressures in a swollen brain. Although the technique is highly controversial, it is often performed as a method of last resort; yet, little is known about how the opening of the skull affects the strain and stress fields inside the brain. Here we introduce a computational model to explore the effects of decompressive craniectomy in idealized and personalized geometries. Through a systematic analysis of different sets of simulations, we identify several common features and trends that could help make the overall procedure less invasive. 

In all cases, a unified stretch pattern with three extreme stretch regions emerges: a tensile zone deep inside the bulge, a highly localized compressive zone around the opening, and a shear zone around the opening. This suggests that regions deep inside the bulge are most vulnerable to damage by axonal stretch, while regions near the craniectomy edge are most vulnerable to damage by axonal shear and herniation. These findings are in agreement with our analytical prediction \cite{goriely16}. Axonal stretch has been studied quite extensively in single axon experiments in vitro \cite{tang10}, but axonal shear has been given little attention as a potential failure mechanism. Only a few studies distinguish between  tension/compression- and shear-type damage \cite{elsayed08}. Our study shows that the tangential stretch, a surrogate measure for the axonal shear, can take values as high as 1.3 for swelling volumes as small as 28\,ml, corresponding to only 5\% of the overall white matter volume. These numbers agree well with a recent simulation that predicted shear strains of the order of 25\% for swelling volumes of 22\,ml \cite{fletcher16}. In view of the long and slender ultrastructure of an axon \cite{bedem15}, it seems fairly reasonable to assume that it could be highly vulnerable to shear-type loading. We have recently shown that shear stresses in bulging solids are highly localized at the bulging edge in regions that we have termed damage drops \cite{weickenmeier16a}. The drop-shaped regions in the tangential stretch profiles of Figure \ref{fig03} agree remarkably well in shape, location, and orientation with our previous analytical predictions \cite{goriely16}.

The recent awareness to shear loading is in line with a current trend in mild traumatic brain injury: For a long time, scientists have thought that linear accelerations are the major origin of brain damage in traumatic brain injury, and that damage would be mainly a result of axonal stretch \cite{rowson12}. Stretch-based injury criteria suggests that there is a 50\% chance of brain tissue damage at strain levels of 18\% in vitro \cite{bain00} and a 50\% chance of mild traumatic brain injury at strain levels of 19\% in vivo \cite{zhang04}. We now know that rotational accelerations could play an equally important role in traumatic brain injury \cite{hernandez15}. With more information becoming available, we might soon recognize that brain damage results not only from stretch \cite{wu16}, but also from shear \cite{elsayed08}, and that the critical damage thresholds need to be considered for each mechanism individually or for both mechanisms in combination \cite{wright11}.  

Our simulations suggest that a frontal craniectomy, which provides anatomic space for a larger opening, creates significantly lower displacements, strains, and stretches in comparison to a unilateral craniectomy \cite{kolias13}. Typically, a craniectomy is performed to release swelling volumes on the order of $50-150$\,ml \cite{fletcher14}. A recent study reported swelling volumes of $27-127$\, ml \cite{holst12}. If we assume that gray and white matter are approximately of the same volume, and that both hemispheres are of equal size, for a total brain volume of 1,108cm$^3$, our swelling of 10\% corresponds to a swelling volume of 56ml for swelling of both hemispheres and to 28ml for unilateral swelling. While our swelling volumes are clearly on the lower end and would probably not be sufficient to require surgery in clinical practice, we already observe significant local strains that could exceed the functional and morphological damage thresholds of 18\% and 21\% reported in the literature \cite{bain00}. 

Our study only presents a first prototype analysis of strains and stretches inside the brain in response to intracranial swelling and decompressive craniectomy. To gain insight, we have made several simplifying assumptions: First, on the time scale of interest, on the order of hours, days, and weeks, we have modeled brain tissue as nonlinearly elastic keeping in mind that on shorter times scales, viscous effects might play an important role \cite{rooij16}. Second, while recent experiments suggest that the elastic response of brain tissues may reasonably well be approximated as isotropic \cite{wright11}, the damage response could very well be anisotropic with different failure mechanisms and different damage thresholds associated with axonal tension and axonal shear \cite{cloots11}. Third, for simplicity, we have assumed that all axons point radially outward. A more realistic model would take into account the discrete axonal orientation at each individual point of the brain \cite{li13}. Conceptually, our analysis itself would remain the same; yet, the post-processing to calculate the normal and shear stretches would use the true axonal direction $\vec{n}$ from diffusion tensor images rather than the simplified  assumption that $\vec{n}$ points radially outward. 
\section{Conclusion}\label{sec:6}
Taken together, our study of bulging brains illustrates how swelling-induced deformations propagate across the brain when opening the skull. It underlines the notion that a decompressive craniectomy is a highly invasive surgical procedure that releases an elevated intracranial pressure at the expense of inducing local zones of extreme strain and stretch. Mathematical models and computational simulations can help identify regions of extreme tissue kinematics. This  approach could guide neurosurgeons to optimize the shape and position of the craniectomy with the goal to avoid placing the craniectomy edge near functionally important regions of the brain. 

\begin{acknowledgements}
\noindent
We thank Allan L. Reiss and his group for providing the MRI scans and Celia Butler and Simpleware for their support in creating the finite element mesh of the brain. 
This work was supported 
by the Timoshenko Scholar Award  to Alain Goriely and
by the Humboldt Research Award and the National Institutes of Health grant U01 HL119578 to Ellen Kuhl. 
\end{acknowledgements}

\end{document}